\RequirePackage[2020-02-02]{latexrelease}
\documentclass[prb,aps, twocolumn, amsmath,amssymb,superscriptaddress]{revtex4}
\usepackage{float}
\usepackage{amsmath}
\usepackage{amssymb}
\usepackage{amsfonts}
\usepackage{euscript}
\usepackage{enumerate}
\usepackage{hhline}
\usepackage{pslatex}
\usepackage{tabularx}
\usepackage[usenames,dvipsnames]{xcolor}
\usepackage[colorlinks,linkcolor=red,anchorcolor=blue,citecolor=blue]{hyperref}

\usepackage{graphicx}
\usepackage{dcolumn}
\usepackage{bm}
\usepackage[sort&compress]{natbib}
\usepackage{balance}
\usepackage{float}
\usepackage{amsmath}
\usepackage{amssymb}
\usepackage{amsfonts}
\usepackage{euscript}
\usepackage{enumerate}
\usepackage{hhline}
\usepackage{pslatex}
\usepackage{tabularx}
\usepackage[usenames,dvipsnames]{xcolor}
\usepackage{graphicx}
\usepackage{dcolumn}
\usepackage{bm}

\usepackage[switch]{lineno}

\makeatletter

\renewcommand{\@biblabel}[1]{#1. }
\renewcommand{\@dotsep}{500}
\renewcommand{\@pnumwidth}{0em}
\renewcommand{\l@figure}[2]{
\@dottedtocline{1}{1.5em}{2em}{Figure #1}{}\vspace{15pt}}

\begin{document}
	
\title{Integrated Optical Vortex Microcomb}

\author{Bo Chen}
\thanks{These authors contributed equally}
\affiliation{State Key Laboratory of Optoelectronic Materials and Technologies, School of Physics, School of Electronics and Information Technology, Sun Yat-sen University, Guangzhou, 510275, China.}

\author{Yueguang Zhou}
\thanks{These authors contributed equally}
\affiliation{DTU Electro, Department of Electrical and Photonics Engineering, Technical University of Denmark, DK-2800 Lyngby, Denmark.}

\author{Yang Liu}
\thanks{These authors contributed equally}
\affiliation{DTU Electro, Department of Electrical and Photonics Engineering, Technical University of Denmark, DK-2800 Lyngby, Denmark.}

\author{Chaochao Ye}
\thanks{These authors contributed equally}
\affiliation{DTU Electro, Department of Electrical and Photonics Engineering, Technical University of Denmark, DK-2800 Lyngby, Denmark.}

\author{Qian Cao}
\thanks{These authors contributed equally}
\affiliation{School of Optical-Electrical and Computer Engineering, University of Shanghai for Science and Technology, Shanghai, China.}
\affiliation{Zhangjiang Laboratory, 100 Haike Road, Shanghai, 201204, China.}

\author{Peinian Huang}
\affiliation{State Key Laboratory of Optoelectronic Materials and Technologies, School of Physics, School of Electronics and Information Technology, Sun Yat-sen University, Guangzhou, 510275, China.}

\author{Chanju Kim}
\affiliation{DTU Electro, Department of Electrical and Photonics Engineering, Technical University of Denmark, DK-2800 Lyngby, Denmark.}

\author{Yi Zheng}
\affiliation{DTU Electro, Department of Electrical and Photonics Engineering, Technical University of Denmark, DK-2800 Lyngby, Denmark.}

\author{Leif Katsuo Oxenløwe}
\affiliation{DTU Electro, Department of Electrical and Photonics Engineering, Technical University of Denmark, DK-2800 Lyngby, Denmark.}

\author{Kresten Yvind}
\affiliation{DTU Electro, Department of Electrical and Photonics Engineering, Technical University of Denmark, DK-2800 Lyngby, Denmark.}

\author{Jin Li}
\affiliation{CAS Key Laboratory of Quantum Information, University of Science and Technology of China, Hefei, 230026, China.}

\author{Jiaqi Li}
\affiliation{State Key Laboratory of Optoelectronic Materials and Technologies, School of Physics, School of Electronics and Information Technology, Sun Yat-sen University, Guangzhou, 510275, China.}

\author{Yanfeng Zhang}
\affiliation{State Key Laboratory of Optoelectronic Materials and Technologies, School of Physics, School of Electronics and Information Technology, Sun Yat-sen University, Guangzhou, 510275, China.}

\author{Chunhua Dong}
\affiliation{CAS Key Laboratory of Quantum Information, University of Science and Technology of China, Hefei, 230026, China.}

\author{Songnian Fu}
\affiliation{Institute of Advanced Photonics Technology, School of Information Engineering, Guangdong University of Technology, Guangzhou, 510006, China.}

\author{Qiwen Zhan}
\thanks{qwzhan@usst.edu.cn}
\affiliation{School of Optical-Electrical and Computer Engineering, University of Shanghai for Science and Technology, Shanghai, China.}
\affiliation{Zhangjiang Laboratory, 100 Haike Road, Shanghai, 201204, China.}
\affiliation{Department of Optical Science and Engineering, Hefei University of Technology, Hefei, Anhui, 230009, China.}

\author{Xuehua Wang}
\thanks{wangxueh@mail.sysu.edu.cn}
\affiliation{State Key Laboratory of Optoelectronic Materials and Technologies, School of Physics, School of Electronics and Information Technology, Sun Yat-sen University, Guangzhou, 510275, China.}
\affiliation{Quantum Science Center of Guangdong-Hong Kong-Macao Greater Bay Area, Shenzhen, China.}

\author{Minhao Pu}
\thanks{mipu@dtu.dk}
\affiliation{DTU Electro, Department of Electrical and Photonics Engineering, Technical University of Denmark, DK-2800 Lyngby, Denmark.}

\author{Jin Liu}
\thanks{liujin23@mail.sysu.edu.cn}
\affiliation{State Key Laboratory of Optoelectronic Materials and Technologies, School of Physics, School of Electronics and Information Technology, Sun Yat-sen University, Guangzhou, 510275, China.}
\affiliation{Quantum Science Center of Guangdong-Hong Kong-Macao Greater Bay Area, Shenzhen, China.}

\date{\today}

\begin{abstract}
\noindent \textbf{The explorations of physical degrees of freedom with infinite dimensionalities, such as orbital angular momentum and frequency of light, have profoundly reshaped the landscape of modern optics with representative photonic functional devices including optical vortex emitters and frequency combs. In nanophotonics, whispering gallery mode microresonators naturally support orbital angular momentum of light and have been demonstrated as on-chip emitters of monochromatic optical vortices. On the other hand, whispering gallery mode microresonators serve as a highly efficient nonlinear optical platform for producing light at different frequencies - i.e., microcombs. Here, we interlace the optical vortices and microcombs by demonstrating an optical vortex comb on an III-V integrated nonlinear microresonator. The angular-grating-dressed nonlinear microring simultaneously emits spatiotemporal light springs consisting of 50 orbital angular momentum modes that are each spectrally addressed to the frequency components (longitudinal whispering gallery modes) of the generated microcomb. We further experimentally generate optical pulses with time-varying orbital angular momenta by carefully introducing a specific intermodal phase relation to spatiotemporal light springs. This work may immediately boost the development of integrated nonlinear/quantum photonics for exploring fundamental optical physics and advancing photonic quantum technology.}
\end{abstract}

\maketitle

Photons with helical phase fronts can carry unlimited, but quantized amount of orbital angular momentum (OAM) \cite{allen1992orbital} and therefore serve as an enabling technology with applications across fundamental physics\cite{he2022towards}, optical communications \cite{wang2012terabit} and quantum photonics \cite{erhard2018twisted}. Circular optical resonators such as microdisks or microrings sustain whispering gallery modes (WGMs) naturally carrying OAM. The introduction of periodic angular gratings to the WGM resonator collectively scatters the WGMs to the free space, leading to on-chip emissions of optical vortices \cite{cai2012integrated}. Compared to other on-chip solutions of vortex beam generators, e.g., metasurfaces \cite{ren2019metasurface,sroor2020high} and fork gratings \cite{zhou2019ultra}, microresonator vortex emitters benefit from strong light-matter interactions due to the prolonged photon lifetimes and enhanced light intensities. For example, spontaneous and stimulated emissions can be dramatically enhanced by the cavity quantum electrodynamics effect, facilitating the realization of highly-efficient OAM single-photon sources \cite{chen2021bright,ma2022chip} and low-threshold vortex microlasers \cite{miao2016orbital,hayenga2019direct,zhang2020tunable}. Hitherto, optical vortex emitters based on either passive or active microrings are operated monochromatically. On the other hand, integrated microresonators in nonlinear materials enable Kerr comb generation (KCG), revolutionizing applications in timekeeping, telecommunication, chemical sensing, distance ranging, etc \cite{gaeta2019photonic,chang2022integrated}. Unlike the OAM emitters, the microcombs rely on high-quality ($ Q $) factor microresonators with an extremely low cavity roundtrip loss \cite{pfeiffer2018ultra,puckett2021422}. Though microresonators support OAM and KCG, the developments of microresonator-based vortex emitters and comb generators have been detached from each other due to their contradicting requirements regarding cavity structure scattering (the former relies on light scattering while the latter necessitates effective suppression of light scattering). Here, we demonstrate a conceptually new class of nanophotonic devices, vortex microcombs, on an integrated AlGaAs photonic platform \cite{pu2016efficient,chang2020ultra}. A delicate resonator design balances the light confinement and vortex emission, where the scattering (emission) loss induced by angular gratings is compensated by the ultra-high nonlinearity of AlGaAs-on-insulator (AlGaAsOI) waveguides \cite{pu2016efficient}. We show that such a vortex microcomb is capable of emitting a massive assembly of spectrally multiplexed optical vortices, each translated from the corresponding spectral component of the microcomb. The unique configuration, featuring varied OAMs distributed across different frequencies, holds the promise of unprecedented advantages in generating spatiotemporal beams with distinctive dynamic features \cite{shen2023roadmap, zhao2020dynamic,chong2020generation, wan2022toroidal, zdagkas2022observation,chen2022synthesizing}. In particular, spatiotemporal light springs with a frequency-OAM correlation and self-torque pulses with time-varying OAMs \cite{rego2019generation,chen2022synthesizing} are accessible from our device with low energy consumption, small device footprint, engineerable OAM configuration and wide spectral coverage.\\

\begin{figure*}[htpb]
	\begin{center}
		\includegraphics[width=0.7\linewidth]{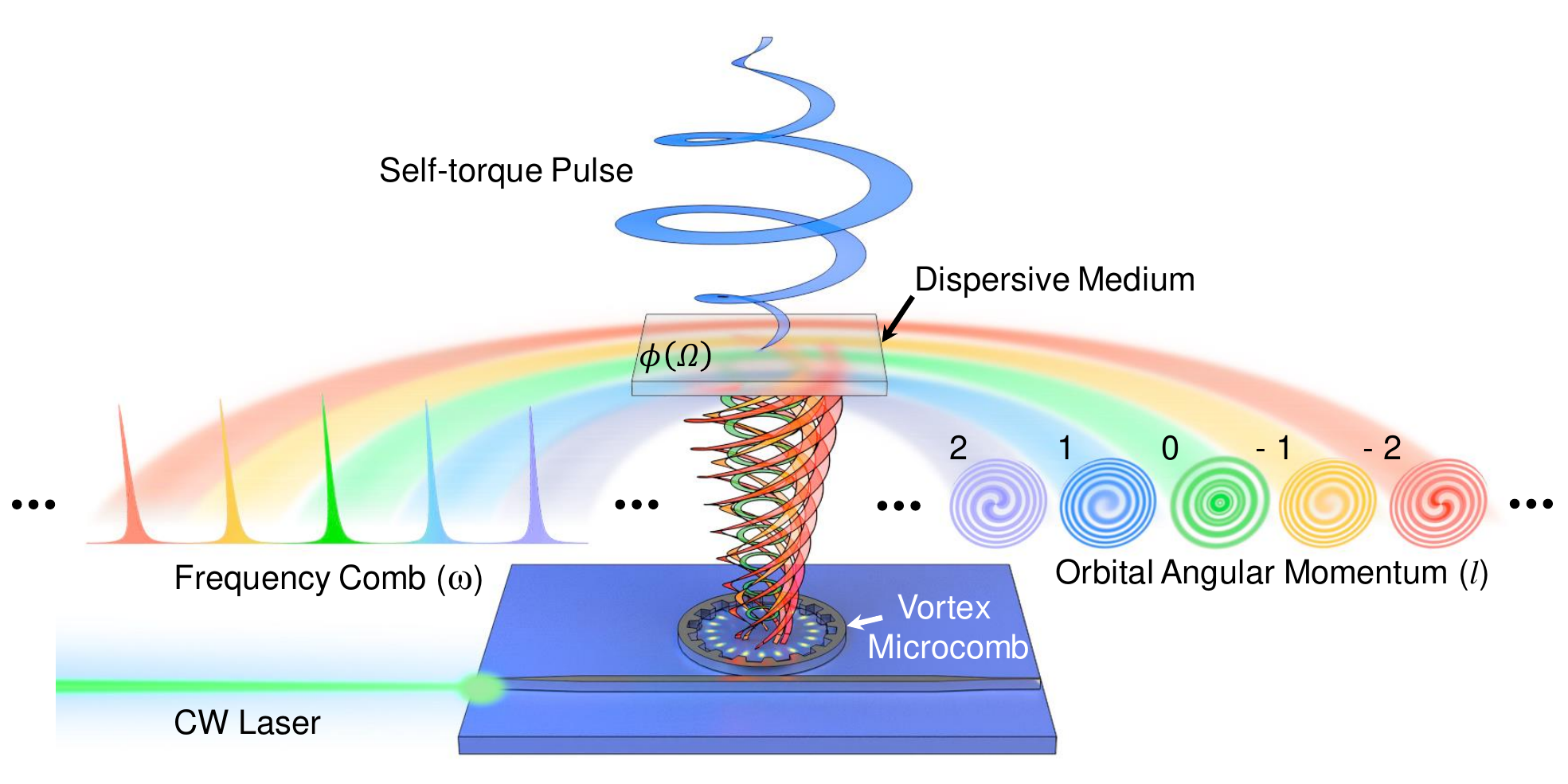}
		\caption{\textbf{Schematic of generating self-torque pulses from the vortex microcomb.} The angular-grating-addressed nonlinear microring emits multiple optical vortices with each OAM mode (right side) in a different frequency (left side). A dispersive medium is employed to generate spatiotemporal optical pulses with varying OAMs.}
		\label{fig:Fig1}
	\end{center}
\end{figure*}

Our devices consist of a high-$ Q $ microring with inner sidewall gratings made from the emerging AlGaAsOI platform exhibiting tight mode confinements and strong optical nonlinearities for low-threshold microcomb generations\cite{pu2016efficient,chang2020ultra}. The vortex microcomb [Fig.~1] is driven by a continuous wave (CW) pump that initializes optical parametric oscillation (OPO) and cascaded four-wave mixing (FWM) between the adjacent WGMs of the microring, resulting in the generation of microcombs with equidistant sidebands. Each comb line corresponds to a WGM with an azimuthal number $ m $, where the angular gratings establish a modal relationship between the emitted optical vortices and the WGMs \cite{cai2012integrated}. The generated comb lines - WGMs - are consequently ejected upwards to the free space, which forms multiple optical vortices with topological charges of $ l= sgn(m)(\lvert m\rvert-N) $ (The sign function $ sgn(m) $ is determined by the counterclockwise (CCW: $ sgn(m) = 1$) and clockwise (CW: $ sgn(m) = -1$) WGM modes, $ N $ is the number of periods of the angular grating). By passing the vortex comb through a dispersive medium, self-torque pulses with time-varying OAMs can be generated, as elaborated in Fig.~(4).\\

The device was fabricated on the AlGaAsOI platform [see Fig.~S1 of SI]. Fig.~2(a) presents scanning electron microscope (SEM) images of a 25-$\mu \rm{m}$-radius microring coupled with a bus waveguide. Fig.~2(b) shows the zoomed coupling region, where the angular gratings can be clearly identified. To mitigate the scattering loss of the fundamental TE$ _{00} $ mode in an OAM microring, we utilize a multi-mode waveguide featuring a cross-sectional dimension of 380~nm $ \times $ 750~nm [see Fig.~S2 of SI]. The mode profile is superimposed in the cross-section SEM image of the microring in Fig.~2(c). The grating protrusion is designed to 30~nm to achieve appreciable efficiencies for vortice ejection and high-$ Q $s for comb generation \cite{lu2023highly}. 274 angular gratings are implemented in the design so that the WGM mode with zero-order OAM falls into the telecom C-band regime. Representative transmission spectra for different resonances of the angular-grating-dressed microring are presented in Fig.~2(d,~e). One can observe a pronounced mode splitting of 9.0~GHz in Fig.~2(d), which is an inherent characteristic feature for the zero-order ($ l=0 $) cavity resonance of the vortex emitter\cite{cai2012integrated,lu2023highly}. For a typical resonance with a non-zero topological charge \cite{puckett2021422} as shown in Fig.~2(e), the mode-splitting is negligible and the extracted intrinsic $ Q $ ($ Q_i $) is $3.1\times10^{5}$.  Fig.~2(f) presents the distributions of $ Q_i $ for the optical vortex emitters and the reference microrings (without angular gratings). A grating size of 30~nm $\times$ 30~nm efficiently scatters WGM modes to the free space while maintaining reasonably high $ Q $ for efficient frequency comb generation via the cavity-enhanced nonlinear optical processes.\\

\begin{figure*}[htpb]
	\begin{center}
		\includegraphics[width=0.7\linewidth]{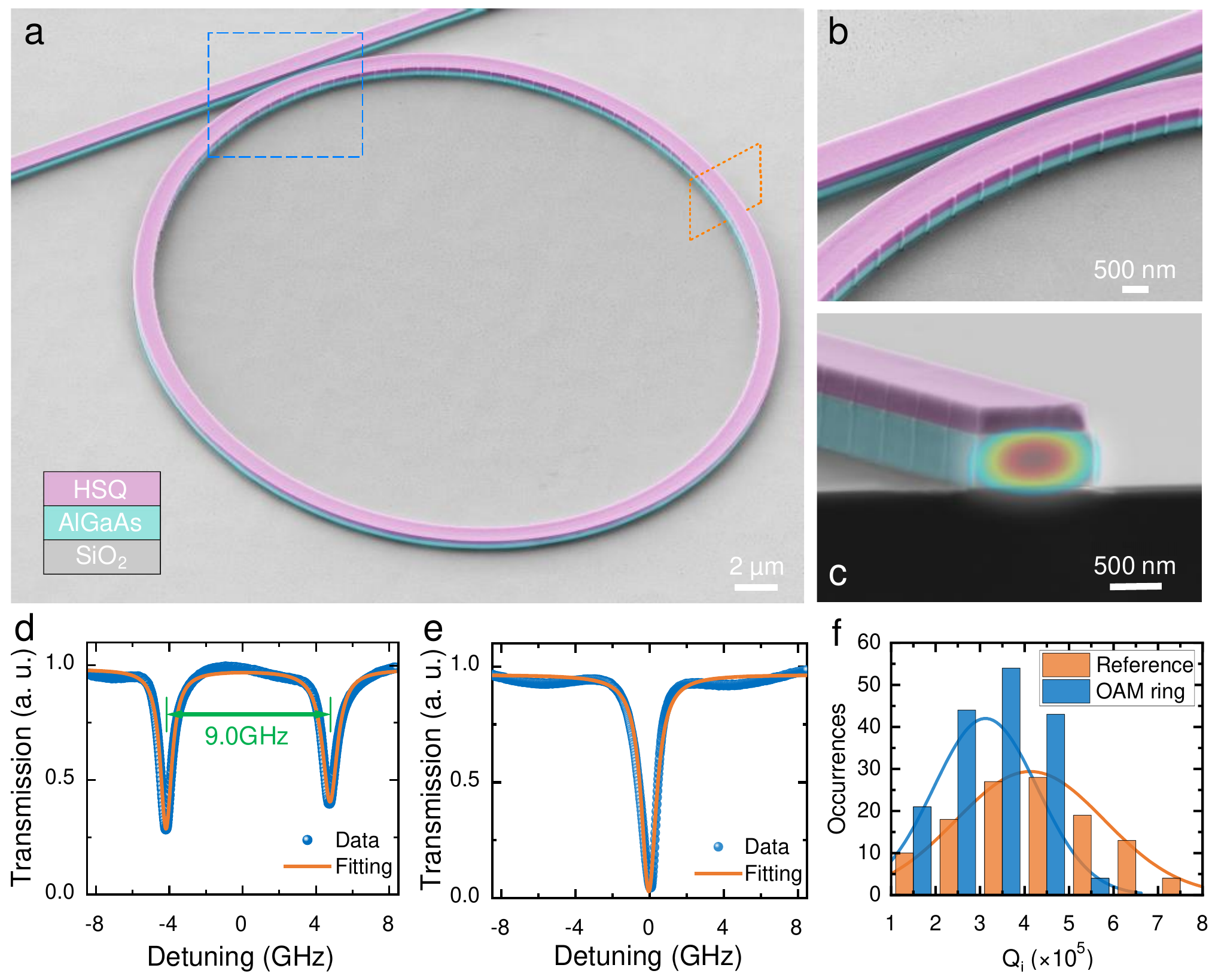}
		\caption{\textbf{Characterizations of the AlGaAsOI microring.} (\textbf{a}) SEM image of the fabricated device. (\textbf{b}) Zoom-in of the blue dashed line highlighted area in (a), showing inner sidewall angular gratings. (\textbf{c}) SEM image of the cross-section (red dashed line highlighted area in (a)) of the microring waveguide, overlaid with the field distribution of the $\rm{TE_{00}}$ mode. (\textbf{d}) Transmission spectrum of the splitted resonance corresponding to zero-order OAM mode with the $ Q_i $ of $3.6 \times 10^5$. (\textbf{e}) Transmission spectrum of a resonance with the median intrinsic $ Q_i $ of $3.1 \times 10^5$,  derived from the statistical analysis of 162 resonances, as shown in Fig. 2(f) . (\textbf{f}) Histogram of  $ Q_i $  distributions for the OAM microring and the reference microring in experimental statistics. The waveguide cross-section dimension is 380$ \times $750 nm$ ^{2} $, and the angular grating feature size of the OAM microring is 30 $\times$ 30~nm$ ^{2} $.  }
		\label{fig:Fig2}
	\end{center}
\end{figure*}

The experimental setup for characterizing the vortex comb is schematically shown in Fig.~3(a). Fig.~3(b) presents the full comb spectrum spanning from 1480 nm to 1680 nm.  All the generated comb lines were ejected to the free space by the embedded angular gratings and further spatially separated by an off-chip transmission grating to resolve different OAM modes. We imaged the far-field patterns of each optical vortex from $ l = 4 $ to $  l= - 4 $ with a high-sensitivity charge-coupled device (CCD) mounted on a movable stage. The direct imaging of the far-field emissions resulted in clockwise (CW) and counterclockwise (CCW) interference patterns [Fig.~3(c)], which can be faithfully reproduced by numerical simulations with an internal dipole source placed in the microring, as shown in Fig.~3(d) [see more discussions in Fig.~S(3,4) of SI]. Such interferometric behaviors have been readily observed in the optical vortices from both low-$ Q $ GaAs \cite{chen2021bright} and high-$ Q $ SiN microrings \cite{lu2023highly}. It has been theoretically and experimentally demonstrated that cylindrical vector vortices from micro-rings with angular gratings can be converted to a right-hand circularly polarized beam with a topological charge of $\lvert l+1\rvert$ and a left-hand circularly polarized beam with a topological charge of $\lvert l-1\rvert$. Therefore, the OAM nature is further confirmed by interfering with the upwards-emitted photons with a reference beam transmitted from the waveguide output, as presented in Fig.~3(e). Before the interference, a quarter waveplate was employed to project vectorial reference beams with topological charge $l$ to the circular polarization basis, forming left-handed polarized beams with $ l’= \lvert l-1\rvert $. The measured interference patterns therefore exhibited spiral arms equal to $\lvert l-1\rvert$, as confirmed by the numerical simulations in Fig.~3(f) (see Fig.~S5 of SI). The projection of the emitted vortices to the right-hand circularly polarized beams with $ l’= \lvert l+1\rvert $ is shown in Fig.~S6 of the SI, showing almost equal composition to the left-hand component. For the vortex beams with large topological charges, e.g., $ l= -5 $ to $ l= – 13 $, the CW and CCW interference patterns exhibit convoluted patterns due to the superpositions of azimuthal and radial components of the cavity field (see Fig.~S7 of SI), which make it challenging to quantitatively compare all the features between experiments and simulations, as presented in Fig.3(g,h). In addition, spiral arms for high-order vortex beams become vague [Fig.~3(e,f)], preventing clear identifications of the topological charges. To reliably extract the high-order topological charges of the vortex beams, we projected the CW/CCW interference patterns to the linear polarization basis, forming clear standing-wave patterns with a number of the anti-modes equal to $ l'=2\lvert l-1\rvert$ [see Fig.~S7 of SI], as presented in Fig.~3(i,j). We use this method to successfully quantify the topological charge up to 25, as shown in Fig.~S8 of SI.

\begin{figure*}[htpb]
	\begin{center}
		\includegraphics[width=0.95\linewidth]{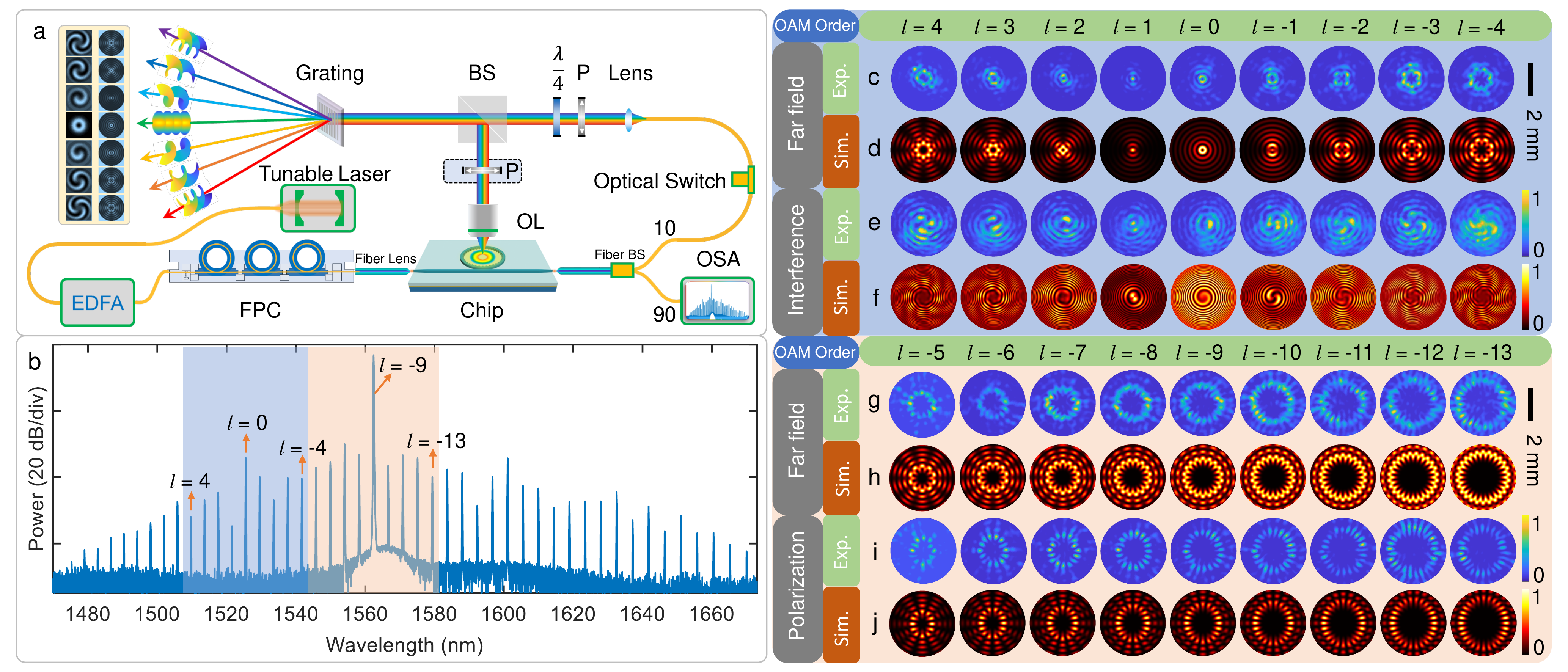}
		\caption{\textbf{Characterizations of the vortex frequency microcomb.} (\textbf{a}) Schematic of the experimental setup. A tunable CW laser is amplified by an EDFA, and its polarization state is controlled by a FPC before coupling to a waveguide via a lensed fiber. The transmitted light from the waveguide output is collected by the other lensed fiber and sent to either an OSA or to the interference path. The ejected optical vortices are either directly imaged on a CCD or combined with a reference beam and then imaged on a CCD after spatial separations by a transmission grating. OL: optical objective. FPC: fiber polarization controller. P: polarizer. $\lambda/4$: quarter-wave plate. EDFA: Erbium-Doped Fiber Amplifier. BS: beam splitter. OSA: optical spectrum analyzer. (\textbf{b}) Spectrum of the microcomb measured by OSA. The light blue area denotes the comb lines with OAM charge from 4 to -4, the light red area represents the comb lines with OAM charge from -5 to -13. (\textbf{c}) Measured and simulated (\textbf{d}) far-field patterns without the reference beam for $ l $ from 4 to -4. (\textbf{e}) Measured and simulated (\textbf{f}) far-field patterns with the reference beam for $ l $ from 4 to -4. (\textbf{g}) Measured and (\textbf{h}) simulated far-field patterns for $ l $ from -5 to -13. (\textbf{i}) Measured and simulated (\textbf{j}) far-field patterns projected to linear polarization basis for $ l $ from -5 to -13.}
		\label{fig:Fig3}
	\end{center}
\end{figure*}

The vortex microcombs may immediately be employed to investigate fundamental optical physics and explore advanced photonic technologies. In particular, the emissions from our devices with an OAM and frequency correlation well aligned with recent breakthroughs in spatiotemporal light spring generations\cite{pariente2015spatio,béjot2021spatiotemporal,piccardo2023broadband}. By further introducing a specific phase relation between different modes, light springs are able to exhibit a time-varying OAM. The self-torque pulse provides unprecedented opportunities to study a self-induced time variation of the
angular momentum in the optical domain, in analogy to the radiation reaction of charged particles \cite{ranada1984on} or gravitational self-fields \cite{dolan2014gravitational}. To the best of our knowledge, the self-torque pulses so far have only been created via high-harmonic generation (HHG) processes \cite{rego2019generation} or by metasurfaces embedded in a Fourier transform setup\cite{chen2022synthesizing}. However, the wavelength of the former case is limited to the extremely ultraviolet regime in HHG processes and the OAM order in the later experiment is rather limited by the metasurfaces. We, now, show, in Fig.~4, self-torque optical pulses can be versatilely synthesized with the vortex microcombs.  
The setup for generating and characterizing the self-torque pulses is schematically shown in Fig.~4(a). The upwards-emitted comb light is propagated through a bandpass filter (remove the pump and select the targeted comb lines for better resolving the varying OAM features) and subsequently enters into a Mach-Zehnder (MZ) interferometer for self-torque pulse generation and characterizations. In one arm of the MZ interferometer, the group delay dispersion (GDD) phase is introduced to the linearly polarized component of the emitted light by a spatial light modulator (SLM) to form optical pulses with time-varying OAM. On the other arm, the comb emissions beam pass through a spatial filter and an optical delay line, serving as a reference beam to interfere with the self-torque pulses for characterizing the dynamic OAM. We note that the spatial filter is deliberately placed in the local field which is linearly polarized with the same orientation as the SLM (see Fig.~S9). Therefore, other polarization components of the emission are effectively filtered out during the interference process due to the polarization mismatch (See more discussions in Fig.~S10 of SI). To maintain a stable phase between different comb lines for the generation of self-torque pulses, we operate the microcomb in a soliton state, as presented in Fig.~4(b). Noise measurements confirm the soliton feature of the microcombs, as shown in Fig.~E1. Further quantitative mode decomposition by employing an SLM is presented in Fig.~E2, showing that the CCW component in the emitted vortices is stronger than their CW counterparts. Three comb lines ($l = 4,~5,~6$) with an intensity variation below 1~dB are filtered to generate self-torque pulses, as highlighted by the red area in Fig.~4(b). The calculated spatiotemporal profile of the targeted self-torque pulse using an iso-surface plot is presented in Fig.~4(c), focusing on the distribution of the main lobe and exhibiting varied OAM at different times in both intensity and phase, as shown in Fig.~4(d). We note that the experimentally extracted mode composition in Fig.~E2 is used in the calculation of Fig.~4(c) (see the comparison in Fig.~E3). The experimentally generated self-torque optical pulses are reconstructed in Fig.~4(e) with their dynamic OAM presented in Fig.~4(f), showing very good agreements with the numerical simulations. We further show in Fig.~4(g) that the microcomb can be operated in a soliton state with a 2-free spectral range (2-FSR) mode interval. By spectrally selecting the comb lines corresponding to $l=6,~8$, a double-helices self-torque beam is obtained as numerically calculated in Fig.~4(h, i) and experimentally demonstrated in Fig.~4(j,k). Compared with the HHG and other approaches, the use of vortex microcombs for generating self-torque space-time pulses offers significant advantages. First, the vortex combs only require a CW laser instead of high-power ultrafast pulse lasers. Second, the topological configurations of the OAM can be versatilely engineered by tuning FSR of the comb states, which is not possible with the HHG approach. Third, the frequency of the self-torque beams from vortex microcombs can be straightforwardly expanded from visible to NIR by employing different material platforms. Finally, our integrated devices are much more compact and suitable for large-scale production using modern nanofabrication technology. 

\begin{figure*}[htpb]
	\begin{center}
		\includegraphics[width=0.85\linewidth]{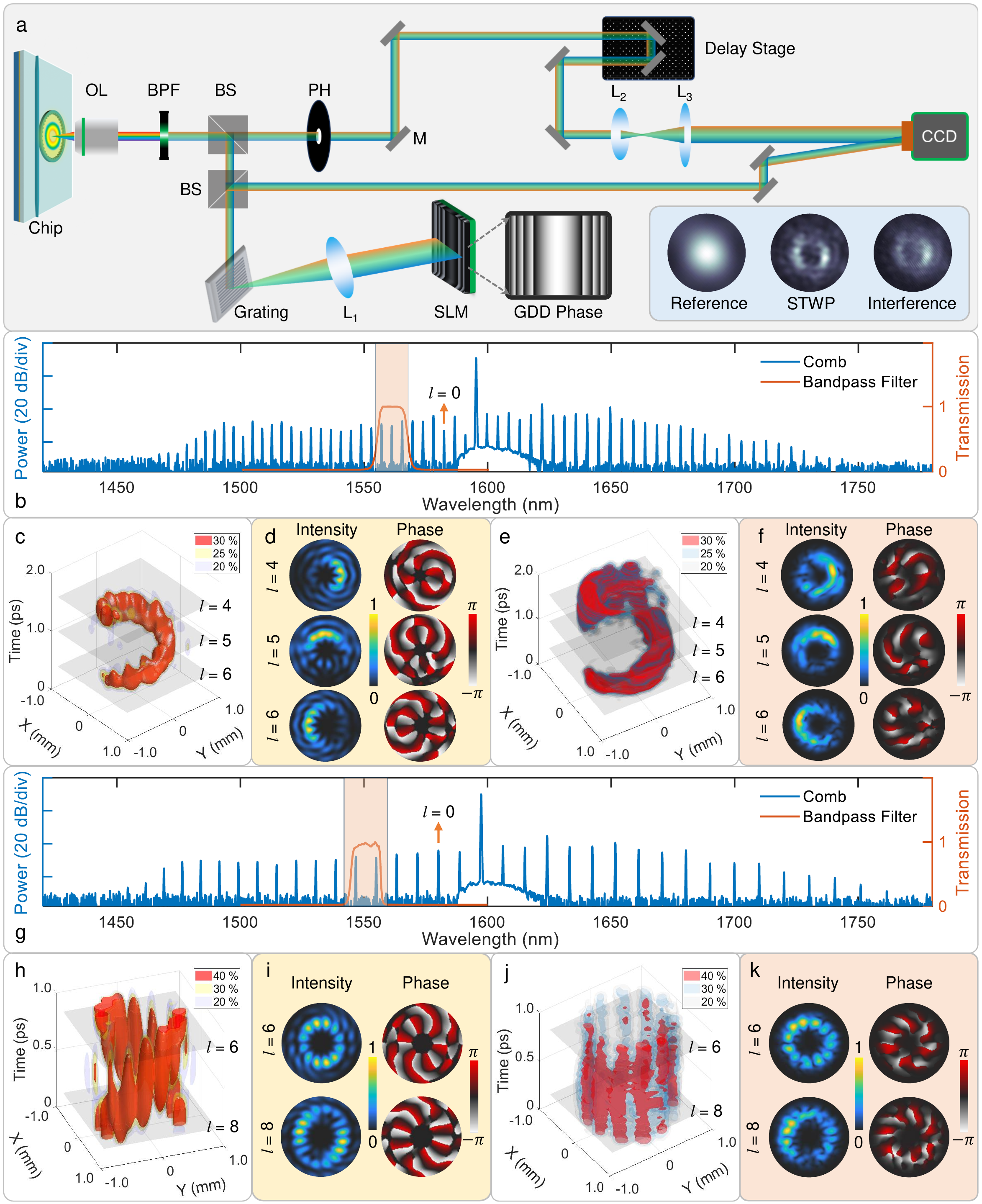}
		\caption{ \textbf{Synthesis of self-torque pulses using vortex microcombs in soliton states.} (\textbf{a}) Experimental setup. A bandpass filter selects the comb lines with targeted frequencies and the filtered light propagates to enter a Mach-Zehnder interferometer with two arms. The upper arm is an optical delay line controlling the reference wavepacket which is shaped into a linearly polarized Gaussian pulse (with the same polarization orientation as the SLM) by the pinhole (see more details in Fig.~S9 of SI). The lower arm introduces a GDD phase to the linearly polarized component of the emitted light for generating the self-torque wavepackets (STWP). The generated self-torque pulses and reference beam overlap in both space and time at the CCD with a small incident angle. By scanning the time delay between the self-torque pulses and the reference beam, the 3D intensity and phase profile of the self-torque wavepacket can be retrieved by analyzing the captured CCD images using the algorithm in Ref.\cite{li2011three}. OL: objective lens, BPF: bandpass filter, PH: pinhole, $L_1 - L_3$: lenses, BS: beam splitter, SLM: spatial light modulator. \textbf{(b), (g)}: Measured spectra of the microcombs in soliton states with different comb mode spacings (500~GHz in (b) and 1000~GHz in (g)). The red lines denote transmissions of the filters spectrally selecting comb lines with OAM charge $l = 4, 5, 6 $ and $l = 6, 8 $, respectively. \textbf{(c), (h)}: Simulated space-time intensity profile of the targeted self-torque pulses by using iso-surface for the wavepacket at 30/25/20\% and 40/30/20\% of its peak intensity generated from the soliton states in (b) and (g), respectively. \textbf{(d), (i)}: Transverse cross-section of the intensity and phase of the self-torque pulses marked by the grey plane in (c) and (h), respectively. \textbf{(e), (j)}: Measured iso-intensity profiles of the self-torque pulses in the corresponding soliton states. \textbf{(f), (k)}: Retrieved intensity and phase from the interference profile of the self-torque pulse marked by the grey plane in (e) and (j), respectively.}
		\label{fig:Fig4}
	\end{center}
\end{figure*}

We have demonstrated a conceptually new nanophotonic device by fusing two separately developing fields, optical vortices and microcombs. Our device can emit up to 50 OAM modes, each translated from corresponding Kerr comb lines generated by an AlGaAs microring. We experimentally demonstrated the generation and engineering of optical pulses with time-varying OAM by operating the microcomb device in different soliton states. Further improvement in the conversion efficiency of the microcomb can be pursued by implementing the recently demonstrated coupled-microresonator design \cite{helgason2023surpassing}. III-V compound semiconductors employed in this work feature both second- and third-order nonlinearities and engineerable bandgaps, which are highly beneficial for integrated nonlinear photonics. Furthermore, the direct bandgap of AlGaAs provides opportunities of monolithic integrations with on-chip lasers and detectors. Moving into the quantum regime, high-dimensional OAM entangled quantum states of light can be created from the vortex microcombs with potential applications in stronger violations of local realism\cite{dada2011experimental, hu2022high} and generating high-dimensional multiphoton entanglement for future quantum technologies. Our method of generating vortex microcombs can be easily implemented in other material platforms \cite{liu2022emerging} such as SiN, SiC, AlN, and LiNbO$_{3}$, etc, providing exciting opportunities to integrated photonics for harnessing structured light-matter interactions.\\

\noindent \textbf{Acknowledgements}
This research is supported by National Key R\&D Program of China (2021YFA1400800), VILLUM FONDEN (VIL50469), European Research Council (REFOCUS 853522), the National Natural Science Foundation of China (62035017, 12334017, 12293052, 12104522, 92050202, 61975243), the Natural Science Foundation of Guangdong (2022A1515011400), Guangdong Introducing Innovative and Entrepreneurial Teams of “The Pearl River Talent Recruitment Program” (2021ZT09X044), the Danish National Research Foundation, SPOC (DNRF123), and Guangdong Special Support Program (2019JC05X397). We thank Yujie Chen, Jie Liu and Siyuan Yu for loaning the equipment.\\

\noindent \textbf{Author contributions}
{J.~L conceived the project. B.~C, Y.~G.~Z, and Q.~C performed the numerical simulations. Y.~G.~Z, C.~C.~Y and Y.~L fabricated the devices. B.~C, P.~N.~H, J.~Li Y.~L C.~C.~Y, J.~Q.~L, Y.~F.~Z and Y.~G.~Z built the setup and characterized the devices. B.~C., P.~N,~H, C.~L, L.~Y, C.~C.~Y, Y.~G.~Z, C.~K, Y.~Z, Q.~C, Q.~W.~Z, M.~H.~P and J.~L analyzed the data. M.~H.~P and J.~L wrote the manuscript with inputs from all authors. C.~H.~D, L.~K.~O, K.~Y, Q.~W.~Z, X.~H.~W, M.~H.~P and J.~L supervised the project.}\\

\noindent \textbf{Competing interests}
{The authors declare no competing financial interests.}\\

\noindent \textbf{Data and materials availability}
{The data sets will be available upon reasonable request.}


\newpage
\noindent \textbf{Methods}

\noindent \textbf{Self-torque pulse:}
The vortex microcomb can emit self-torque optical pulses with a time-varying OAM. According to ref.\cite{zhu2014spin}, the spatial profile of these comb modes in the far-field can be written as
\begin{equation}
\mathrm{E} = \mathrm{E}_{\rho} + e^{i\pi/2} \mathrm{E}_{\varphi}
\end{equation}
\begin{equation}
    \mathrm{E}_{\rho} \propto (-i)^{Nl}{\frac{1}{\sqrt{\rho^{2}+\zeta^{2}}}} e^{iNl\varphi}\left( J_{Nl+1} +J_{Nl-1} \right) 
\end{equation}
\begin{equation}
    \mathrm{E}_{\varphi} \propto (-i)^{N(l+1)}{\frac{1}{\sqrt{\rho^{2}+\zeta^{2}}}} e^{iNl\varphi}\left( J_{Nl+1} -J_{Nl-1}\right) 
\end{equation}
	where $ l $ is the topological charge of the comb mode and N is the harmonic order of the mode-locking state, $J_{n} = J_{n}(\nu\rho/\zeta)$ denotes the $n^{th}$ order Bessel function of the first kind. In order to generate self-torque optical pulses, the vortex microcomb is phase modulated in the spectral domain. The applied spectral phase mask can be written as
	\begin{equation}
	\phi(\Omega) = \tau_0 \lvert \Omega \rvert+\mathrm{GDD}\cdot\Omega^2
	\end{equation}
	where $  \Omega = \omega - \omega_0 $, $ \tau_0 $ is the linear phase coefficient, and GDD is the group delay dispersion (GDD) coefficient. such a phase modulation could be implemented experimentally by using a dispersive medium or a programmable pulse shaper. In the experiment, the GDD phase is introduced by a spatial light modulator (SLM) to form optical pulses with time-varying OAM.

\newpage
\onecolumngrid \bigskip

\begin{center} {{\bf \large EXTENDED DATA}}\end{center}

\setcounter{figure}{0}
\makeatletter
\renewcommand{\thefigure}{E\@arabic\c@figure}
\begin{center}
	\begin{figure}[!h]
		\includegraphics[width=0.8\linewidth]{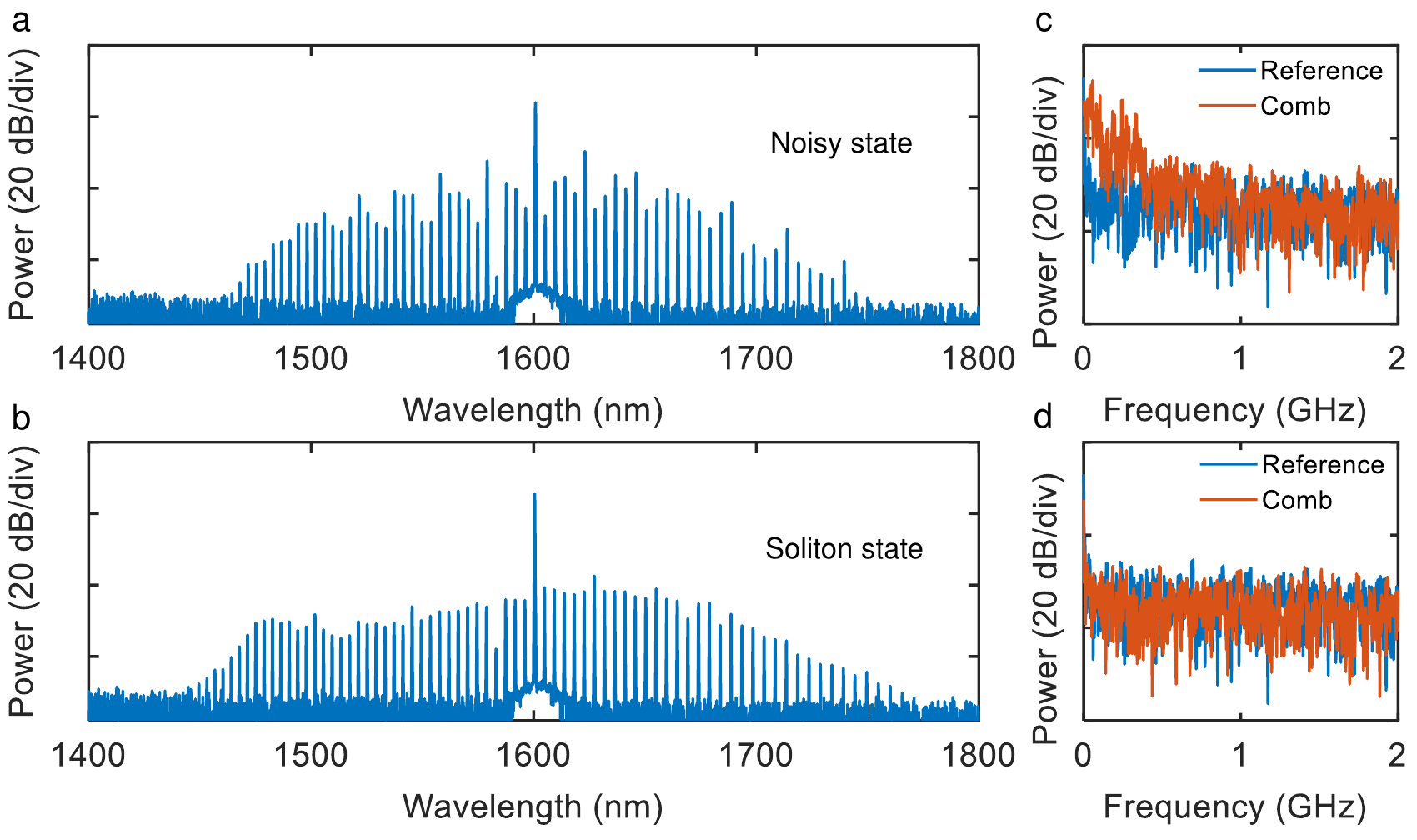}
		\caption{\textbf{Noise characterizations for microcomb states.} \textbf{(a, b)} show the representative microcomb spectra in a noisy state (a) and a soliton state (b). \textbf{(c, d)} show the corresponding RF spectra for different comb states (red). The blue traces show the reference traces recorded with no comb being generated. The increased power below 1 GHz (the red trace in (c)) suggests a noisy comb state.}
		\label{fig:Fig5}
	\end{figure}
\end{center}

\newpage

\begin{center}
	\begin{figure}[!h]
		\includegraphics[width=1\linewidth]{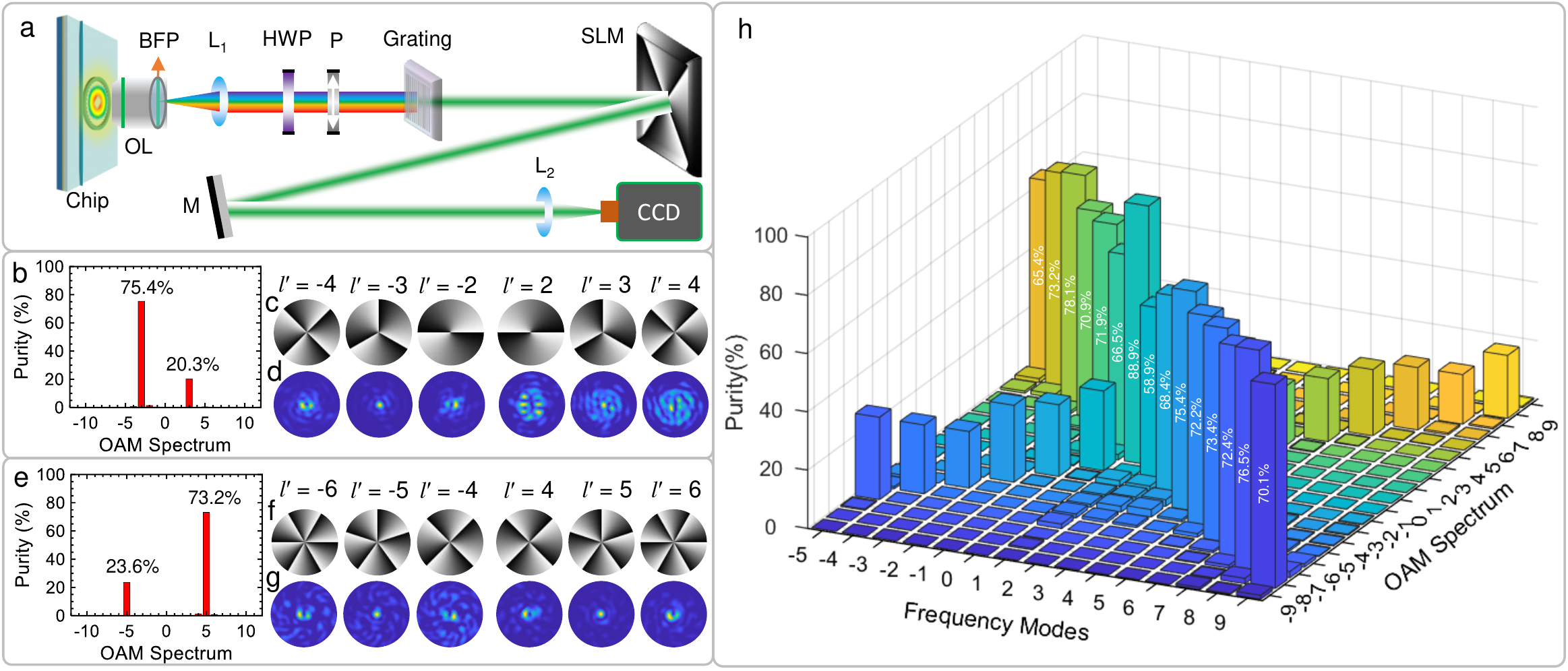}
		\caption{\textbf{Mode decomposition of the vortex microcomb.}(\textbf{a}) Schematic of the experimental setup for the OAM mode purity measurement. OL: the objective lens, BFP: the back focal plane, HWP: half-wave plate, P: polarizer, SLM: spatial light modulator, M: mirror, L$_1$ - L$_2$: lenses, CCD: charge-coupled device.  (\textbf{b, e}) Measured on-axis intensity distributions for the linearly polarized component of the emitted superposition mode with $ l $ = 4 and $ l $ = -4. (\textbf{c, f}) Phase distributions applied to the SLM. (\textbf{d, g}) Measured far-field patterns for the linearly polarized component of the superposition mode. (\textbf{h}) OAM spectrum for frequency OAM modes from $ l $ = -5 to $ l $  = 9.}
		\label{fig:Fig5}
	\end{figure}
\end{center}

\newpage
\begin{center}
	\begin{figure}[!h]
		\includegraphics[width=0.9\linewidth]{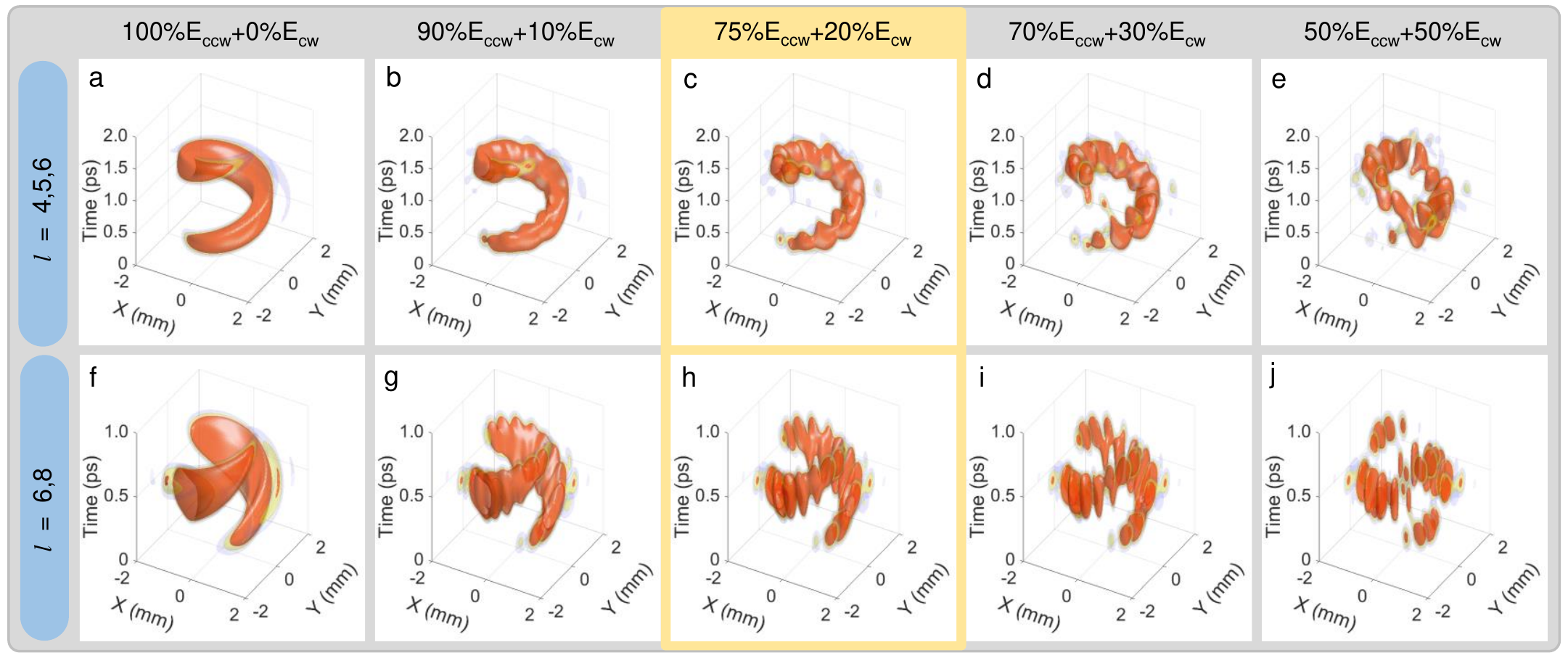}
		\caption{\textbf{Simulation of the self-torque pules with the different CW/CCW compositions.} (\textbf{a-e}) Simulations for the single-helices self-torque pulses. (\textbf{f-j}) Simulations for the double-helices pulses. (c) and (h) are the results presented in Fig.~4 which employes the values extracted from the mode decomposition measurement in Fig.~E2.}
		\label{fig:Fig5}
	\end{figure}
\end{center}

\end{document}